\documentclass[conference]{IEEEtran}
\usepackage{graphicx}
\usepackage{cite}
\usepackage{amsmath}
\usepackage{amsfonts}
\usepackage{amssymb}
\usepackage{algorithm}
\usepackage{algorithmic}
\usepackage{subcaption}
\usepackage{psfrag}
\usepackage{epsfig}
\usepackage{multicol}
\usepackage{tikz}

\usetikzlibrary{positioning}
\usetikzlibrary{shapes,arrows}

 \usetikzlibrary{patterns} 
\usetikzlibrary{decorations.pathreplacing}

\newcommand{\qed}{\nobreak \ifvmode \relax \else
      \ifdim\lastskip<1.5em \hskip-\lastskip
      \hskip1.5em plus0em minus0.5em \fi \nobreak
      \vrule height0.75em width0.5em depth0.25em\fi}

\newcommand{\bm}{\mathbf}
\newcommand{\be}{\begin{equation}}
\newcommand{\ee}{\end{equation}}
\newcommand{\bse}{\begin{subequations}}
\newcommand{\ese}{\end{subequations}}
\newcommand{\bea}{\begin{eqnarray}}
\newcommand{\eea}{\end{eqnarray}}
\newcommand{\x}{{\bm x}}
\newcommand{\y}{{\bm y}}

\newcommand{\ba}{{\bm a}}
\newcommand{\bn}{{\bm n}}

\newcommand{\br}{{\bm r}}

\newcommand{\bA}{{\bm A}}

\newcommand{\bI}{{\bm I}}
\newcommand{\bR}{{\bm R}}

\newcommand{\bE}{{\bf E}}
\newcommand{\bF}{{\bf F}}

\newcommand{\bC}{{\bf C}}
\newcommand{\bB}{{\bf B}}
\newcommand{\bS}{{\bf S}}
\newcommand{\bG}{{\bf G}}
\newcommand{\bH}{{\bf H}}
\newcommand{\bX}{{\bf X}}

\newcommand{\h}{{\bf h}}

\newcommand{\bh}{{\bf h}}

\newcommand{\bs}{{\bf s}}

\newcommand{\bU}{{\bf U}}
\newcommand{\by}{{\bf y}}

\newcommand{\bzero}{{\bf 0}}

\newcommand{\BH}{{\boldsymbol{\mathcal H}}}
\newcommand{\Bh}{{\boldsymbol{ h}}}

\newcommand{\bGamma}{\mbox{\boldmath$\Gamma$}}

\newcommand{\bPhi}{\mbox{\boldmath{$\Phi$}}}

\makeatletter
\newcommand\fs@spaceruled{\def\@fs@cfont{\bfseries}\let\@fs@capt\floatc@ruled
  \def\@fs@pre{\vspace{0.5\baselineskip}\hrule height.8pt depth0pt \kern2pt}%
  \def\@fs@post{\kern2pt\hrule\relax}%
  \def\@fs@mid{\kern2pt\hrule\kern2pt}%
  \let\@fs@iftopcapt\iftrue}
\makeatother
\begin{document}

\raggedbottom

\title{Channel Estimation using 5G Sounding Reference Signals: A Delay-Doppler Domain Approach}
\author{\normalsize Danilo Lelin Li$^*$, $\text{Ramtin Rabiee}^{\dagger}$, Arman Farhang$^*$ \\ $ ^*$Department of Electronic \& Electrical Engineering, Trinity College Dublin, Ireland \\
$ ^\dagger$Huawei Technologies Sweden AB, Stockholm, Sweden\\
Email: \{dlelinli, arman.farhang\}@tcd.ie, ramtin.rabiee@huawei.com
\vspace{-0.4cm} } \vspace{-0.7cm}

\maketitle

\begin{abstract}

Delay-Doppler multicarrier modulation (DDMC) techniques have been among the central topics of research for high-Doppler channels. However, a complete transition to DDMC-based waveforms is not yet practically feasible. This is because 5G NR based waveforms, orthogonal frequency division multiplexing (OFDM) and discrete Fourier transform-spread OFDM (DFT-s-OFDM), remain as the modulation schemes for the sixth-generation radio (6GR). Hence, in this paper, we demonstrate how we can still benefit from DD-domain processing in high-mobility scenarios using 5G NR sounding reference signals (SRSs). By considering a DFT-s-OFDM receiver, we transform each received OFDM symbol into the delay-Doppler (DD) domain, where the channel is then estimated. With this approach, we estimate the DD channel parameters, allowing us to predict the aged channel over OFDM symbols without pilots. To improve channel prediction, we propose a linear joint channel estimation and equalization technique, where we use the detected data in each OFDM symbol to sequentially update our channel estimates.
Our simulation results show that the proposed technique significantly outperforms the conventional frequency-domain estimation technique in terms of bit error rate (BER) and normalized mean squared error (NMSE). Furthermore, we show that using only two slots with SRS for initial channel estimation, our method supports pilot-free detection for more than 25 subsequent OFDM symbols.

\end{abstract}


\section{Introduction}
\label{sec:intro}
Orthogonal frequency division multiplexing (OFDM) has served as the waveform of choice for the 4th and 5th generations of wireless networks and remains as the dominant air-interface technology in 6G.
Its main advantage lies in its robustness to multipath fading in linear time-invariant (LTI) channels where each subcarrier experiences a flat-fading channel.
This enables efficient channel estimation and low complexity single-tap equalization. However, future 6G applications and use cases such as autonomous driving (aerial and terrestrial), unmanned aerial vehicles (UAVs), high-speed rail, low-Earth-orbit (LEO) satellites and millimeter-wave links will encounter doubly selective channels. These channels severely degrade OFDM performance and represent one of the key challenges in 6G networks \cite{CientificChallenges6G}. Consequently, significant research is being directed toward alternative waveform designs.

Delay-Doppler (DD) domain signal processing has attracted significant attention in recent years due to its advantages in linear time-varying (LTV) channel conditions \cite{CodedOTFS,DDaided,MP1,MP2,BEM2,SC-FDMA,Embedded,ReducedCP}. Modulation schemes such as Orthogonal Time Frequency Space (OTFS) have demonstrated lower bit error rates (BER) and improved spectral efficiency compared to OFDM in high-mobility scenarios \cite{CodedOTFS}. However, as much as the 6G requirements envision revolutionary technology, it also requires an evolutionary approach. A complete replacement of OFDM with OTFS remains infeasible in near-term deployments. Hence, a DD-aided OFDM framework was proposed in \cite{DDaided}, where the time-frequency domain channel estimates are transformed into the DD domain to enable more precise estimation of the channel parameters using maximum likelihood or peak detection-based estimators. 
However, such techniques and other estimators in the literature on DD domain processing, \cite{MP1,MP2}, are characterized by high computational complexity. 
Alternative low-complexity estimators include those that deploy the basis expansion model (BEM) to estimate channel parameters on a grid \cite{BEM1,BEM2}. However, accuracy of these solutions largely depend on the number of available channel estimates, which incurs large pilot overhead. Additionally, DD domain processing techniques often experience latency issues. They require receiving a large number of delay blocks, each corresponding to an OFDM symbol, before the channel can be reliably estimated in the DD domain.

Considering these limitations, this paper proposes an alternative framework that benefits from DD-domain processing in high-mobility scenarios. Furthermore, it ensures compatibility with current 5G NR standard and employs linear channel estimation and equalization techniques. 
In our proposed framework, the transmitter sends 5G~NR time–frequency domain reference signals. The receiver then estimates the DD channel using a DFT-spread OFDM (DFT-s-OFDM) demodulator, which maps each OFDM symbol onto a DD grid.
This is possible as DFT-s-OFDM can be seen as a DD modulation scheme from the perspective of the DFT-precoders input and the inverse DFT (IDFT) post-processing unit outputs at the transmitter and receiver, respectively \cite{SC-FDMA}.
This framework does not suffer from latency issues, as each DD grid is generated immediately after receiving each OFDM symbol, rather than waiting for multiple symbols. To estimate the channel, we consider sounding reference signals (SRSs), where pilot and data signals are allocated to different OFDM symbols, see Fig.~\ref{Fig:Slotstructure}. 
We take the SRS pilot symbols to the DD domain and linearly estimate the channel. Then, the DD domain channel is transformed into the delay-time domain which is used as an input to the BEM to estimate the on-grid DD channel parameters. This allows us to predict the channel over the transmitted OFDM data symbols.

It is important to note that, according to 5G NR standard \cite{3gpp_ts38211_v1860}, SRS is generally allocated within the final OFDM symbols of each slot. 
Consequently, this allocation strategy necessitates the reception of the entire slot before reliable channel estimation can be performed. This preserves the latency challenges associated with waiting for complete slot reception.
To mitigate this issue, we propose transmitting only two slots that include SRS for initial channel estimation, while subsequent slots carry data exclusively. To improve the channel prediction over long data sequences without pilots, we propose using the detected data of each OFDM symbol, and sequentially update the channel parameters before detecting the next received symbol. Through simulations, we show that our proposed DD framework significantly improves the normalized mean squared error (NMSE) and BER performance when compared to time-frequency channel estimation. Furthermore, the proposed sequential channel estimator and equalizer using the detected data further improves BER performance and allows for transmission of multiple slots containing only data signals, after an initial stream of SRS slots.

{\it Notations}: 
$[\bA]_{m,n}$
represents the element in the $m^{\rm{th}}$ row and $n^{\rm{th}}$ column of matrix $\bA$. $\bI_M$ and $\textbf{0}_{M\times N}$ are the identity and zero matrices of the sizes $M \times M$ and $M \times N$, respectively. circ$\{ \ba \}$ and diag$\{ \ba \}$ denote a circulant matrix with the first column formed by vector $\ba$ and a diagonal matrix with the diagonal elements  formed by $\ba$,  respectively. {CircShift$\bigl( \bA,[r, c]\bigr)$ represents an operator that circularly shifts the elements of matrix $\bA$ by $r$ rows and $c$ columns.} The superscripts $(\cdot)^{\rm T}$ and $(\cdot)^{\rm H}$ indicate transpose and conjugate transpose operations, respectively.
$|\cdot |_M$, $\otimes$ and $\odot$ denote modulo $M$, Kronecker product and Hadamard product operators, respectively.
$\delta(\cdot)$ is the Dirac delta function. Finally, $\bF_M$ is the normalized $M$-point DFT matrix.

\section{System Model}
\label{sec:sysmod}

\begin{figure}

	\centering
	\usetikzlibrary{fit}
\usetikzlibrary{positioning}
\usetikzlibrary{shapes,arrows}
 \usetikzlibrary{patterns} 
\usetikzlibrary{decorations.pathreplacing}

\begin{tikzpicture}
\def\a{-2};
\def\c{0};
\def\h{0.2};
\def\w{0.4};

\draw[fill=green!30](\a+\w,\c+\h)rectangle (\a+11*\w,\c+15*\h);
\draw[fill=orange!60](\a,\c+\h)rectangle (\a+\w,\c+15*\h);
\draw[fill=orange!60](\a,\c)rectangle (\a+15*\w,\c+\h);
\foreach \x in {11,...,14}{
\foreach \y in {0,...,3}{
\draw[fill=blue!60](\a+\x*\w,\c+4*\y*\h+\h)rectangle (\a+\x*\w+\w,\c+4*\y*\h+2*\h);
    }
}


\draw[xstep=0.4 cm,ystep=0.2 cm,color=black] (\a,\c) grid (\a+15*\w,\c+15*\h);

\def\x{0};\def\y{5};

\def\y{8};
\node at (\a+11*\w+0.125,\c+\y*\h+0.125)(a1){};

\def\x{0};\def\y{0};
\foreach \x in {1,...,14}
\node at (\a+\x*\w+0.125,\c+\y*\h+0.11){\tiny \x};

\def\x{0};\def\y{0};
\foreach \y in {1,...,14}
\node at (\a+\x*\w+0.2,\c+\y*\h+0.11){\tiny \y};

\node at (\a+7*\w+0.2,\c+16*\h+0.11){\small $\vdots$};

\node[rotate=90] at (\a-0.25,\c+10*\h+0.125){{\footnotesize \rm{Frequency}}};
\node at (\a+7*\w+0.15,\c-0.25){{\footnotesize \rm{Time}}};

\draw[fill=green!30](\a+16*\w,\c)rectangle (\a+16*\w+0.25,\c+0.25)node[xshift=0.8 cm,yshift=-0.15 cm,align=center]{{\small Data signal}};
\def\c{0.5};
\draw[fill=blue!60](\a+16*\w,\c)rectangle (\a+16*\w+0.25,\c+0.25)node[xshift=0.8 cm,yshift=-0.15 cm]{{\small Pilot signal}};

\end{tikzpicture}
    \caption{SRS slot structure.}
    \vspace{-0.5cm}
\label{Fig:Slotstructure}

\end{figure}
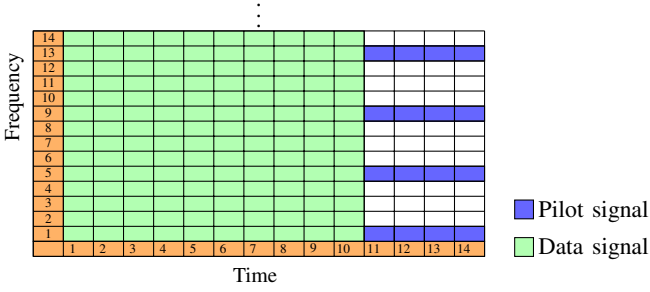

We consider a time-frequency slot transmitting Physical Uplink Shared Channel (PUSCH) data, with $N_{\rm{o}}=14$ OFDM symbols each containing $M_{\rm{o}}$ subcarriers with a spacing of $\Delta f$. This results in a bandwidth of $B_{w}=M_{\rm{o}} \Delta f$ and a sampling period of $T_s=\frac{1}{B_w}$. {In accordance with 3GPP 5GNR standard \cite{3gpp_ts38211_v1860}, the slot transmits only PUSCH data in OFDM symbols 1-10, and only SRS in the last four OFDM symbols using a comb-4 pattern, i.e., $K_{\rm{TC}}=4$, that spans along the whole frequency band}, as shown in Fig.~\ref{Fig:Slotstructure}. Let the $M_{\rm{o}} \times N_{\rm{o}}$ matrix $\bX$ represent the transmitted slot, comprised of data signals $\bX^{(\rm{d})}$ and reference signals $\bX^{(\rm{p})}$, i.e., $\bX=\bX^{(\rm{d})}+\bX^{(\rm{p})}$. The time domain transmit signal $\bS$ of size $M_{\rm{T}} \times N_{\rm{o}}$ (where $M_{\rm{T}}=M_{\rm{o}}+M_{\rm{CP}}$) is obtained with an $M_{\rm{o}}$-point IDFT, after which a cyclic prefix (CP) of length $M_{\rm{CP}}$ larger than the channel length is appended at the beginning of each OFDM symbol. The OFDM modulation can be summarized as
\be
    \bS = \bA_{\rm{CP}} \bF^{\rm{H}}_{M_{\rm{o}}}\bX,
\ee
where the CP is added using $\bA_{\rm{CP}} = [\bG_{\rm{CP}}^{\rm T} , \bI_{M_{\rm{o}}} ]^{\rm T}$ of size $M_{\rm{T}} \times M_{\rm{o}}$, and $\bG_{\rm{CP}}$ includes the last $M_{\rm{CP}}$ rows of $\bI_{M_{\rm{o}}}$. 

The signal then goes through parallel-to-serial conversion of $\bS$ to obtain $\bs= [\bs_0^{\rm T},\ldots,\bs_{N_{\rm{o}}-1}^{\rm T}]^{\rm T}$, where $\bs_{n_{{\rm{o}}}}$ represents the $n$-th column of $\bS$. The resulting signal is transmitted over an LTV channel, and the received signal can be written as 
\begin{align}
    \begin{split}\label{eq:linch}
        r[k] 
        =\sum_{\ell=0}^{L-1}  [\mathcal{{{H}}}]_{\ell,k} s[k-\ell]+ \eta[k],
    \end{split}
\end{align}
where vector $\boldsymbol{\eta}$ includes the complex additive white gaussian noise (AWGN), with the variance $\sigma^2$, i.e., $\eta[k] \sim \mathcal{CN}(0,\sigma^2)$. The channel contains $P$ paths and $L$ delay taps and its delay-time domain response is represented by the $L\times M_{\rm{T}}N_{\rm{o}} $ matrix $\BH$, with its elements given by
\be\label{eq:h}
[\mathcal{{{H}}}]_{\ell,k} = \sum_{p=0}^{P-1} \alpha_{p} e^{j2\pi \upsilon_{p} k T_{\rm{s}}} \delta [\ell-\ell_{\tau_{p}}],
\ee
where the parameters $\alpha_{p}$, $\tau_{p}$ and $\upsilon_{p}$ represent the path gains, delays and Doppler shifts of path $p$, respectively.

To pave the way for the derivations in the following sections, the received signal can be represented in the vectorized form $\br = \bH \bs + \boldsymbol{\eta}$, where $\bH$ represents the channel matrix of size $M_{\rm{T}}N_{\rm{o}} \times M_{\rm{T}}N_{\rm{o}} $ and its elements are expressed as 
\be
[\bH]_{k,\ell}=\begin{cases}
    [\mathcal{H}]_{k-\ell,k} &\text{ if } k-\ell \in [0,L-1], \\
    0 &\text{ if } k-\ell \notin [0,L-1].
\end{cases}
\ee

To represent each received OFDM symbol separately, we partition the matrix $\bH$ into an $N_{\rm{o}} \times N_{\rm{o}}$ grid, where each block is an $M_{\rm{T}} \times M_{\rm{T}}$ submatrix. We denote by $\bH^{(u,v)}$ the submatrix located in the $u$‑th row and $v$‑th column of this block partitioning, with $u,v \in [0,\dots,N_{\rm{o}}-1]$. 
After CP removal, the received signal on the $n$-th OFDM symbol of length $M_{\rm{o}}$ can be represented as
\be
\br_{n_{{\rm{o}}}} = \bR_{\rm{CP}} \bH^{(n_{\rm{o}},n_{\rm{o}})} \bs_{n_{{\rm{o}}}}+\boldsymbol{\eta}_{n_{{\rm{o}}}}=\bH_{n_{{\rm{o}}}} \bF^{\rm{H}}_{M_{\rm{o}}} \x_{n_{{\rm{o}}}} +\boldsymbol{\eta}_{n_{{\rm{o}}}},
\ee
where $\x_{n_{{\rm{o}}}} $ represents the $n_{{\rm{o}}}$-th column of $\bX$, $\bR_{\rm{CP}}=[\bzero_{M_{\rm{o}} \times M_{\rm{CP}}},\bI_{M_{\rm{o}}}]$ represents the CP removal matrix, and the elements of $\bH_{n_{{\rm{o}}}}=\bR_{\rm{CP}} \bH^{(n_{\rm{o}},n_{\rm{o}})} \bA_{\rm{CP}}$ are given as 
\be
[\bH_{n_{{\rm{o}}}}]_{k,\ell}=\begin{cases}
    [\mathcal{H}]_{|k-\ell|_{M_{\rm{o}}},k+M_{\rm{CP}}+n_{\rm{o}}M_{\rm{T}}}&\text{ if } |k-\ell|_{M_{\rm{o}}} < L, \\
    0 &\text{ if } |k-\ell|_{M_{\rm{o}}} \geq L.
\end{cases}
\ee

To retrieve the transmitted signal, the received signal is demodulated with an $M_{\rm{o}}$-point DFT matrix
\be\label{eq:rfreq}
    \by_{n_{{\rm{o}}}} = \bF_{M_{\rm{o}}}\bH_{n_{{\rm{o}}}} \bF^{\rm{H}}_{M_{\rm{o}}} \x_{n_{{\rm{o}}}} +\widetilde{\boldsymbol{\eta}}_{n_{{\rm{o}}}}= \bH^{\rm{OFDM}}_{n_{{\rm{o}}}}\x_{n_{{\rm{o}}}} +\widetilde{\boldsymbol{\eta}}_{n_{{\rm{o}}}},
\ee
where $\widetilde{\boldsymbol{\eta}}_{n_{{\rm{o}}}}$ represents the frequency domain AWGN and 
\be\label{eq:HeqOFDM}
\bH^{\rm{OFDM}}_{n_{{\rm{o}}}}=\bF_{M_{\rm{o}}}\bH_{n_{{\rm{o}}}} \bF^{\rm{H}}_{M_{\rm{o}}}
\ee
represents the equivalent channel at the $n$-th OFDM symbol.

\subsection{Time-Frequency to DD Domain}
\label{sec:RS}

We transform each OFDM symbol into a DD grid with $N$ delay blocks of size $M$, i.e., $M_{\rm{o}}=MN$. Each delay block can be obtained  by first downsampling $\by_{n_{{\rm{o}}}}$ by a factor of $N$, then applaying an $M$-point IDFT, followed by multiplication with a phase compensation matrix. More specifically, the $n$-th delay block from the $n_{\rm{o}}$-th OFDM symbol can be obtained as
\be\label{eq:nthDDdomainr}
    \by^{(n_{{\rm{o}}})}_n = {\rm{diag}}\{ \boldsymbol{\omega}_n \} \bF_{M}^{\rm{H}} \boldsymbol{\Psi}_n \by_{n_{{\rm{o}}}},
\ee
where 
\be
\begin{split}
    \boldsymbol{\Psi}_n&=\text{CircShift}\bigl( ( \bI_M \otimes [1,\bzero_{1,N-1}]),[0, n]\bigr), \text{ and }\\
    \boldsymbol{\omega}_n &= [e^{j\frac{2 \pi 0 n }{MN}},e^{j\frac{2 \pi 1 n }{MN}}, \dots , e^{j\frac{2 \pi (M-1) n }{MN}}  ]
\end{split}
\ee
denotes the downsampling operation matrix and the phase compensating vector, respectively. 
Hence, the vectorized form of the DD domain grid from the $n_{\rm{o}}$-th OFDM symbol $ \by^{(n_{{\rm{o}}})}=[{\by^{(n_{{\rm{o}}})}_0}^{\rm{T}}, \dots, {\by^{(n_{{\rm{o}}})}_{N-1}}^{\rm{T}}]^{\rm{T}}$ can be expressed as 
\be\label{eq:vecDDdomainr}
    \by^{(n_{{\rm{o}}})}=  \boldsymbol{\Omega}  (\bI_N \otimes \bF_{M}^{\rm{H}}) \boldsymbol{\Psi} \bF_{M_{\rm{o}}}\br_{n_{{\rm{o}}}},
\ee
where $\boldsymbol{\Psi} = [\boldsymbol{\Psi}_0^{\rm{T}},\dots, \boldsymbol{\Psi}_{N-1}^{\rm{T}}]^{\rm{T}} $, and $\boldsymbol{\Omega}={\rm{diag}}\{ [\boldsymbol{\omega}_0^{\rm{T}},\dots , \boldsymbol{\omega}_{N-1}^{\rm{T}} ]^{\rm{T}} \}$.

{Observe that if we omit the phase compensating matrix $\boldsymbol{\Omega}$ from \eqref{eq:vecDDdomainr}, the demodulation structure corresponds to an interleaved SC-FDMA demodulator. Hence, transforming the received signal to the DD domain can be achieved by adding a phase compensation matrix to a DFT-s-OFDM scheme, which is envisioned for 6G. Moreover, it follows that}
\be\label{eq:SC-FDMA=OTFS}
    \boldsymbol{\Omega}  (\bI_N \otimes \bF_{M}^{\rm{H}}) \boldsymbol{\Psi} \bF_{M_{\rm{o}}} = (\bF_N \otimes \bI_M  ).
\ee
{In other words, this confirms that \eqref{eq:vecDDdomainr} is equivalent to an OTFS demodulator represented on the right-hand side of \eqref{eq:SC-FDMA=OTFS}. }

{From \eqref{eq:nthDDdomainr} we observe that if $N$ is chosen to be equal to the reference signal spacing, the pilot signal will be contained within a single delay block. Furthermore, if the reference signal sequence is generated by repeating the same value, it becomes an impulsive pilot in the DD domain, the most common pilot scheme in OTFS literature \cite{Embedded}.  }
However, the reference signal sequence in 5G is generated using Zadoff Chu (ZC) or pseudo random sequences. Therefore, in the following section, we study how to estimate the channel in the DD domain using 5G standard pilots.

\section{Channel Estimation in DD Domain}
\label{sec:DDchest}

By estimating the channel in the DD domain, we can improve the accuracy of the estimates by increasing the frequency at which these estimates are obtained. Rather than assuming the channel to be LTI over the entire duration of an OFDM symbol, we instead approximate the channel to be LTI within each smaller delay block. Consequently, instead of representing the channel matrix  $\bH_{n_{\rm{o}}}$ with a single CIR, we characterize it using a set of CIR vectors across each delay block, denoted by $\BH_{n_{\rm{o}}}=[\Bh_{n_{\rm{o}}M_{\rm{T}}+M_{\rm{CP}}},\dots , \Bh_{n_{{\rm{o}}}M_{\rm{T}}+M_{\rm{CP}}+(N-1)M}]$.

To estimate the DD domain channel, the BS assumes that the pilot signal was transmitted in the DD domain.  Let 
\be\label{eq:pilotinDD}
\overline{\x}^{(n_{\rm{o}})}=\boldsymbol{\Omega}  (\bI_N \otimes \bF_{M}^{\rm{H}}) \boldsymbol{\Psi}{\x}_{n_{{\rm{o}}}}
\ee
represent the $n_{\rm{o}}$-th transmitted OFDM symbol transformed to the DD domain. Considering DD transmit signal in \eqref{eq:pilotinDD}, the received signal in the DD domain in \eqref{eq:vecDDdomainr} can be written as
\be\label{eq:vecDDdomainr2}
   \by^{(n_{{\rm{o}}})}= \bH^{\rm{DD}}_{n_{{\rm{o}}}} \overline{\x}^{(n_{\rm{o}})} + {\boldsymbol{\eta}}^{\rm{DD}}_{n_{{\rm{o}}}}, 
\ee
where ${\boldsymbol{\eta}}^{\rm{DD}}_{n_{{\rm{o}}}}$ represents the DD domain AWGN, and the $M_{\rm{o}} \times M_{\rm{o}}$ matrix $\bH^{\rm{DD}}_{n_{{\rm{o}}}}$ represents the equivalent DD domain channel, given as
\be
    \bH^{\rm{DD}}_{n_{{\rm{o}}}}=
    (\bF_N \otimes \bI_M  )  \bH_{n_{{\rm{o}}}}  (\bF_N^{\rm{H}} \otimes \bI_M  ).
\ee

In DD domain modulation literature, the channel is often approximated as LTI within each delay block, with a CP appended at the beginning of every block. This modeling transforms the equivalent DD channel into a doubly block-circulant matrix, which greatly simplifies channel estimation and equalization. In contrast, in our framework each OFDM symbol is mapped onto a different DD grid, and a CP is inserted only at the start of each grid, making the structure similar to that in \cite{ReducedCP}.
With a single CP appended at the start of the DD grid, $\bH^{\rm{DD}}_{n_{{\rm{o}}}}$ assumes the following structure
\be\label{eq:HDDstructure}
   \bH^{\rm{DD}}_{n_{{\rm{o}}}}= \begin{bmatrix}
    \bB_{0,0}^{(n_{\rm{o}})} & \bB_{N-1,1}^{(n_{\rm{o}})}& \dots& \bB^{(n_{\rm{o}})}_{1,N-1}\\
    \bB^{(n_{\rm{o}})}_{1,0} & \bB^{(n_{\rm{o}})}_{0,1}  & \dots& \bB^{(n_{\rm{o}})}_{2,N-1}\\
    \vdots & \vdots& \ddots& \vdots\\
    \bB^{(n_{\rm{o}})}_{N-1,0} & \bB^{(n_{\rm{o}})}_{N-2,1} & \dots &\bB^{(n_{\rm{o}})}_{0,N-1}
    \end{bmatrix},
\ee
where the $M\times M$ matrix $\bB^{(n_{\rm{o}})}_{n,0}= \text{circ} \{ [\widetilde{\Bh}^{\rm{T}}_{n_{\rm{o}},n},\bzero_{1\times  M-L}]^{\rm{T}}  \}$, and $\widetilde{\Bh}_{n_{\rm{o}},n}$ denotes the $n$th column of $\widetilde{\BH}_{n_{\rm{o}}}=\BH_{n_{\rm{o}}}\bF_N$. The $L\times N$ matrix  $\widetilde{\BH}_{n_{\rm{o}}}$ represents the DD impulse response. The submatrix $\bB^{(n_{\rm{o}})}_{n,n'}$ is obtained from $\bB^{(n_{\rm{o}})}_{n,0}$ with all entries above the main diagonal multiplied by the phase factor $e^{2j\pi \frac{n'}{N}}$, i.e.,
\be
    \bB^{(n_{\rm{o}})}_{n,n'} =  \text{circ} \{ [\widetilde{\Bh}^{\rm{T}}_{n_{\rm{o}},n},\bzero_{1\times  M-L}]^{\rm{T}}  \} \odot \bU_{n'},
\ee
where $\bU_{n'}$ is an $M \times M$ matrix with the elements given by
\be
    [\bU_{n'}]_{m,m'} =
    \begin{cases}
        e^{\frac{2j\pi n'}{N}}, & \text{if } m < m' \\
        1, & \text{if } m \geq m'.
    \end{cases}
\ee

This structure withholds the property of doubly circulant matrices, where the doubly circulant matrix and vector can be interchanged, i.e.,    $\bH^{\rm{DD}}_{n_{\rm{o}}} \overline{\x}^{(n_{\rm{o}})} =
    \bGamma_{n_{\rm{o}}} {\bh}^{\rm{DD}}_{n_{\rm{o}}}$, where ${\bh}^{\rm{DD}}_{n_{\rm{o}}}=[[\widetilde{\Bh}_{n_{\rm{o}},0}^{\rm{T}}, \bzero_{1\times M-L}],\dots, [\widetilde{\Bh}_{n_{\rm{o}},N-1}^{\rm{T}}, \bzero_{1\times M-L}]]^{\rm{T}}$ denotes the first column of $\bH^{\rm{DD}}_{n_{\rm{o}}}$, and
\be\label{eq:gammastructure}
   \bGamma_{n_{\rm{o}}}= \begin{bmatrix}
    \bC^{(n_{\rm{o}})}_{0,0} & \bC^{(n_{\rm{o}})}_{N-1,1}& \dots& \bC^{(n_{\rm{o}})}_{1,N-1}\\
    \bC^{(n_{\rm{o}})}_{1,0} & \bC^{(n_{\rm{o}})}_{0,1}  & \dots& \bC^{(n_{\rm{o}})}_{2,N-1}\\
    \vdots & \vdots& \ddots& \vdots\\
    \bC^{(n_{\rm{o}})}_{N-1,0} & \bC^{(n_{\rm{o}})}_{N-2,1} & \dots &\bC^{(n_{\rm{o}})}_{0,N-1}
    \end{bmatrix}
\ee
follows the same structure as $\bH^{\rm{DD}}_{n_{\rm{o}}}$, with 
\be
    \bC^{(n_{\rm{o}})}_{n,n'} =  \text{circ} \{ \overline{\x}^{(n_{{\rm{o}}})}_{n}  \} \odot \bU_{n'}.
\ee

Given that $\bh^{\rm{DD}}_{n_{\rm{o}}}$ is constructed by appending $M-L$ zeros at the end of each DD channel vector $\widetilde{\Bh}_{n_{\rm{o}},n}^{\rm{T}}$, we can eliminate the columns of $\bGamma_{n_{\rm{o}}}$ that corresponds to these zero entries, i.e., removing the last $M-L$ columns of each matrix $\bC^{(n_{\rm{o}})}_{n,n'}$ in \eqref{eq:gammastructure}. The resulting matrix after this column elimination is a $M_{\rm{o}} \times NL$ matrix denoted as $\widetilde{\bGamma}_{n_{\rm{o}}}$, and maintains the equality $\bH^{\rm{DD}}_{n_{\rm{o}}} \overline{\x}^{(n_{\rm{o}})} = \bGamma_{n_{\rm{o}}} {\bh}^{\rm{DD}}_{n_{\rm{o}}}=\widetilde{\bGamma}_{n_{\rm{o}}}\widetilde{\Bh}_{n_{\rm{o}}}$, where $\widetilde{\Bh}_{n_{\rm{o}}}=[\widetilde{\Bh}_{n_{\rm{o}},0}^{\rm{T}},\dots, \widetilde{\Bh}_{n_{\rm{o}},N-1}^{\rm{T}}]^{\rm{T}}$. This way, substituting $\widetilde{\bGamma}_{n_{\rm{o}}}\widetilde{\Bh}_{n_{\rm{o}}}$ in \eqref{eq:vecDDdomainr2}, we have 
\be\label{eq:vecDDdomainr3}
\by^{(n_{{\rm{o}}})}=\widetilde{\bGamma}_{n_{\rm{o}}}\widetilde{\Bh}_{n_{\rm{o}}} +  {\boldsymbol{\eta}}^{\rm{DD}}_{n_{{\rm{o}}}}.
\ee
The DD channel can then be estimated as 
\be\label{eq:estDDdatainterf}
\Hat{\Bh}_{n_{\rm{o}}}^{\rm{DD}}= {\widetilde{\bGamma}_{n_{\rm{o}}}}^{\dagger}\by^{(n_{{\rm{o}}})}=\widetilde{\Bh}_{n_{\rm{o}}} +  {\widetilde{\bGamma}_{n_{\rm{o}}}}^{\dagger} {\boldsymbol{\eta}}^{\rm{DD}}_{n_{{\rm{o}}}}.
\ee

The DD domain channel estimates are then transformed into the time domain to obtain $\widehat{\BH}_{n_{\rm{o}}}$, where the $N$ CIR estimates are interpolated to obtain an improved estimation of $\bH_{n_{{\rm{o}}}}$. Hence, the DD domain channel estimation offers an improved estimation of $\bH_{n_{{\rm{o}}}}$, as opposed to estimating it in the frequency domain, where $\bH_{n_{{\rm{o}}}}$ is constructed using a single CIR $\Bh_{n_{\rm{o}}}$.

{It is important to mention that the design of  DD grid parameters requires careful study. 
When the number of delay blocks equals the pilot subcarrier interval $N=K_{\mathrm{TC}}$, the DD domain pilot occupies a single delay block. In contrast, larger values of $N$ cause the pilot to span multiple delay blocks. Designing for a larger number of delay blocks leads to more channel estimates. However, it also affects the conditioning of matrix $\widetilde{\bGamma}_{n_{\rm{o}}}$. Given the standard ZC sequence for pilot sequence generation, designing $N$ to be larger than $K_{\mathrm{TC}}$ leads to an ill-conditioned matrix $\widetilde{\bGamma}_{n_{\rm{o}}}$. }

\section{Channel Prediction}
\label{sec:pred}

One may observe from the 5G slot structures in Fig.~\ref{Fig:Slotstructure} that most of the OFDM symbols do not contain pilot. Moreover, in the SRS slot structure, all transmitted data precedes the pilot signals. Therefore, simple interpolation of the channel estimates becomes insufficient, and channel prediction is required. In this section, we propose estimating the delay and Doppler parameters by approximating the fractional Doppler shifts into a more granular on grid Doppler taps using the BEM. By obtaining the DD parameters, it becomes possible to characterize the channel within the whole slot through \eqref{eq:h}.

To estimate the channel parameters, we first approximate the Doppler shifts of each channel path $p$ to be within $Q+1$ grid points, i.e., $\upsilon_p \in [0, \pm \frac{2\upsilon_{\rm{max}}}{Q},\pm \frac{4\upsilon_{\rm{max}}}{Q},\dots, \pm \upsilon_{\rm{max}} ] $, where $\upsilon_{\rm{max}}$ denotes the maximum Doppler shift of the channel. With this approximation, \eqref{eq:h} can be rewritten as
\be
\begin{split}\label{eq:GridDopRep}
     [\mathcal{H}]_{\ell,k}=&\sum_{q=-Q/2}^{Q/2} e^{\frac{4j\pi \upsilon_{\rm{max}} q k T_s}{Q}} \alpha_{q,\ell},
\end{split}
\ee
where
\be
\alpha_{q,\ell}= \sum^{P-1}_{p=0} \alpha_{p} \delta[\ell-\ell_{\tau_p}] \delta[\upsilon_p - \frac{2\upsilon_{\rm{max}} q}{Q}].
\ee
The summation in \eqref{eq:GridDopRep} can then be transformed into a matrix multiplication as
\be
    \BH= {\bA}\boldsymbol{\Phi}_{\rm{Full}},
\ee
where  the $L \times Q+1 $ matrix  ${\bA}$ representing the channel gains at each DD pair is defined as
\be
    {\bA}=\begin{bmatrix}
        \alpha_{-Q/2,0}&\dots & \alpha_{Q/2,0}\\
        \vdots & \ddots &\vdots\\
        \alpha_{-Q/2,L-1}&\dots & \alpha_{Q/2,L-1}
    \end{bmatrix}
\ee
and the $Q+1 \times M_{\rm{T}}N_{\rm{o}} $ matrix that maps the Doppler shifts into the time domain is expressed as
\be
    \boldsymbol{\Phi}_{\rm{Full}}=\begin{bmatrix}
        e^{2j\pi \upsilon_{\rm{-Q/2,l}} (0) T_s} & \dots & e^{2j\pi \upsilon_{\rm{-Q/2,l}} (M_{\rm{T}}N_{\rm{o}}-1) T_s} \\
        \vdots & \ddots &\vdots\\
        e^{2j\pi \upsilon_{\rm{Q/2,l}} (0) T_s} & \dots & e^{2j\pi \upsilon_{\rm{Q/2,l}} (M_{\rm{T}}N_{\rm{o}}-1) T_s}
    \end{bmatrix}.  
\ee

Notice from \eqref{eq:estDDdatainterf} that in each OFDM symbol that contains pilots, we estimate $N$ CIRs. Therefore, we can stack the estimated CIRs for all OFDM symbols containing pilots to form $\widehat{\BH}_{\rm{DT}}$, and all the columns in $\bPhi_{\rm{Full}}$ corresponding to the CIR positions in the time domain to form matrix  $\bPhi_{\rm{RS}}$  to obtain the expression
\be\label{eq:curlyHdtest}
    \widehat{\BH}_{\rm{DT}}= \bA \bPhi_{\rm{RS}}.
\ee
Hence, channel gains at each DD pair can  be estimated as
\be\label{eq:estA}
    \widehat{\bA} = \widehat{\BH}_{\rm{DT}} \bPhi_{\rm{RS}}^{\dagger}.
\ee

Finally, from \eqref{eq:estA} the full channel response in the delay-time domain can be estimated as
\be\label{eq:finalest}
\widehat{\BH} = \widehat{\bA}\boldsymbol{\Phi}_{\rm{Full}}.
\ee

\subsection{MMSE-based Channel Equalization}

To detect the transmitted signal, we consider the Minimum Mean Square Error (MMSE) equalizer. The data in the $n_{\rm{o}}$-th OFDM symbol is equalized with the following steps:
\begin{itemize}
    \item Based on the estimated delay-time channel response in \eqref{eq:finalest}, we reconstruct $\widehat{\bH}_{n_{{\rm{o}}}}$, and subsequently using \eqref{eq:HeqOFDM}, the equivalent frequency domain channel matrix $\widehat{\bH}^{\rm{OFDM}}_{n_{{\rm{o}}}}$.
    \item Obtained $\widehat{\bH}^{\rm{OFDM}}_{n_{{\rm{o}}}}$, we form MMSE equalizer matrix as 
    \be
        \bE\! =\! \big( (\widehat{\bH}^{\rm{OFDM}}_{n_{{\rm{o}}}})^{\rm{H}}\widehat{\bH}^{\rm{OFDM}}_{n_{{\rm{o}}}}\! + \bI_{M_{\rm{o}}} \sigma^2  \big)^{-1} (\widehat{\bH}^{\rm{OFDM}}_{n_{{\rm{o}}}})^{\rm{H}}.
    \ee
    \item Finally, the transmitted signal at the $n_{\rm{o}}$-th OFDM symbol is estimated as
    \be\label{eq:MMSExhat}
       \widehat{\x}_{n_{{\rm{o}}}} = \bE \by_{n_{{\rm{o}}}}.
    \ee
\end{itemize}

\subsection{Proposed Data-driven Channel Estimation and Equalization Technique}

While the channel prediction method discussed in this section can detect data symbols transmitted over OFDM symbols without pilot subcarriers, its performance still degrades under high-Doppler scenarios due to rapid channel variations. To address this limitation, we propose a data-driven approach in which the detected data from each OFDM symbol is utilized to refine the channel estimates before detecting the data of subsequent OFDM symbols.

Consider a scenario where multiple slots are transmitted where only the first two slots contain SRS signals. An initial channel estimate is obtained using the SRS from these slots based on equations \eqref{eq:curlyHdtest}–\eqref{eq:finalest}. In this initial state, since 8 SRS OFDM symbols are transmitted, $8N$ CIRs are estimated, resulting in a $\widehat{\BH}_{\rm{DT}}$ matrix of dimension $L \times 8N$. We then estimate the DD-domain channel gains $\widehat{\bA}$ and { define the vector $\bn_{\rm{order}}$ containing the data symbol indices ordered from closest to farthest from the SRS symbols.}
Using this ordering, the initial channel estimate equalizes data from OFDM symbol $n_{\rm{order}}[1]$ using \eqref{eq:MMSExhat}. After detection of the first OFDM symbol, the hard-detected data are employed as virtual pilots to generate updated channel estimates using \eqref{eq:estDDdatainterf}. This process yields $N$ additional CIRs, which are appended to $\widehat{\BH}_{\rm{DT}}$, forming an updated matrix of size $L \times 9N$ and enabling improved estimation of the DD-domain channel gains. The procedure is then repeated after each OFDM symbol detection, progressively increasing the number of estimated CIRs and refining the channel estimates before equalizing the next OFDM symbol. The proposed technique is summarized in Algorithm~\ref{alg:Propsedtechnique}.

\floatstyle{spaceruled}
\restylefloat{algorithm}

\begin{algorithm}[t]
\vspace{-0.05in}

\begin{algorithmic}[1]

\caption{Proposed data-driven channel estimation and equalization technique.}
\label{alg:Propsedtechnique}

\STATE Estimate the initial $\widehat{\BH}_{\rm{DT}}$ with available reference signal. 
\STATE Estimate the channel gain at each DD pair $\widehat{\bA}$ using \eqref{eq:estA}.
\STATE Define the order in which the OFDM data symbols will be equalized $\bn_{\rm{order}}$.

\FOR{$n_{\rm{o}}$ in $\bn_{\rm{order}}$}
    \STATE Obtain $\widehat{\bH}^{\rm{OFDM}}_{n_{{\rm{o}}}}$ using the most recent estimate of $\widehat{\bA}$.
    \STATE Detect the data $\widehat{\x}_{n_{{\rm{o}}}}$ using \eqref{eq:MMSExhat} and apply hard decision.
    \STATE After applying hard decision to $\widehat{\x}_{n_{{\rm{o}}}}$, transform it from time-frequency domain to the DD domain using \eqref{eq:pilotinDD}.
    \STATE Generate $\widetilde{\bGamma}_{n_{\rm{o}}}$, using \eqref{eq:gammastructure} and the detected data in DD domain from \eqref{eq:pilotinDD}, estimate the additional CIRs using \eqref{eq:estDDdatainterf}, and append them to $\widehat{\BH}_{\rm{DT}}$.
    \STATE Update the channel gain at each DD pair $\widehat{\bA}$ using the new $\widehat{\BH}_{\rm{DT}}$ and $\bPhi$, according to $n_{\rm{o}}$.

\ENDFOR

\end{algorithmic}
\end{algorithm}

This method substantially enhances channel estimation performance, enabling reliable data detection for OFDM symbols located far from the reference signal transmission. Moreover, the proposed approach introduces only a minor increase in computational complexity, arising from the inversion of the data matrix $\widetilde{\bGamma}_{n_{\rm{o}}}$, while the inversion of the matrix $\bPhi$ for each OFDM symbol can be performed offline. It is worth noting that the use of SRS introduces additional latency to the system, as it is positioned at the end of each slot. Consequently, it is necessary to wait until the last OFDM symbol of the final slot containing SRS is received before performing data detection. However, the proposed method mitigates this latency by enabling the transmission of slots without SRS. Moreover, the received OFDM symbols following the SRS are processed sequentially, rather than waiting for the entire transmission to complete before performing estimation and equalization, as is typically required in OTFS systems.

\section{Numerical Results}
\label{sec:numres}

\begin{figure}
    \centering
     
        \includegraphics[width=0.9\linewidth]{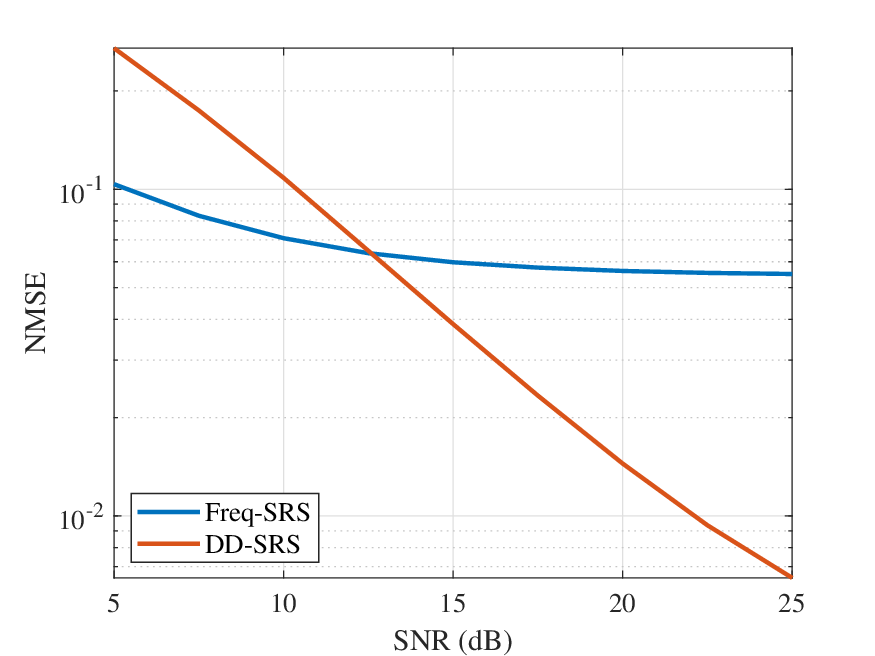}
    \caption{NMSE as a function of SNR for v$=500$~km/h.}
    \label{fig:SingleOFDMsymbolNMSE}

\end{figure}

\begin{figure}[t!]
    \centering
        \includegraphics[width=0.9\linewidth]{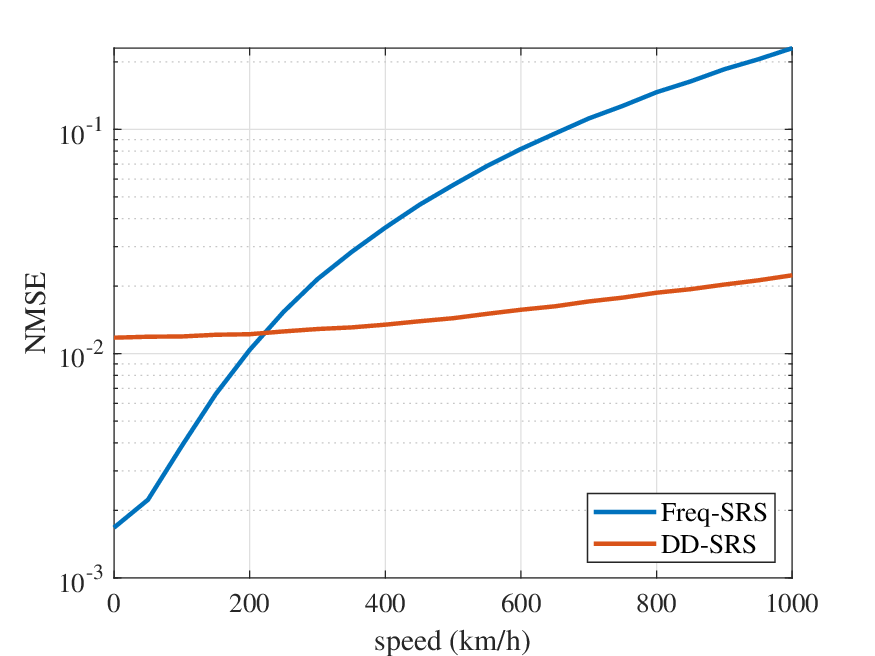}
    \caption{NMSE as a function of relative speed at SNR of 20~dB. }
    \label{fig:SingleOFDMsymbolNMSExspeed}

\end{figure}

In this section, we numerically analyze the efficacy of our proposed framework. We consider $M_{\rm{o}}=1024$ subcarriers per OFDM symbol that transforms into a DD grid with $N=4$ delay blocks of size $M=256$. The transmitted slot structure is shown in Fig.~\ref{Fig:Slotstructure}. In our simulation setup, we consider the tap delay line C model (TDL-C) from 5G standards \cite{3gppTR38901}, which leads to a maximum delay of $L=40$. The Doppler shift is generated using Jake's model with a carrier frequency of $f_{\rm c}=4.9$~GHz and subcarrier spacing of $\Delta f = 15$~kHz. The data is modulated with 4-QAM, while the pilot sequence in each OFDM symbol is generated with a ZC sequence. To avoid ISI, we append a CP of length $M_{\rm{CP}}=39$.

\begin{figure}
    \centering
    \includegraphics[width=0.9\linewidth]{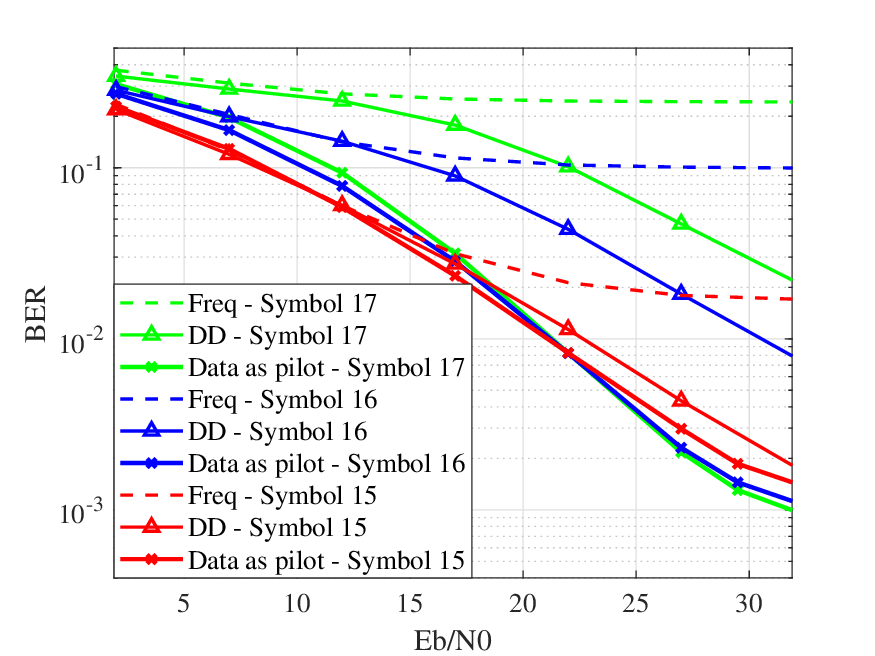}
    \caption{BER comparison between frequency-domain and DD domain channel estimation of data symbols without pilots.}
    \label{fig:BER}
    \vspace{-0.5cm}
\end{figure}

In Fig.~\ref{fig:SingleOFDMsymbolNMSE} and Fig.~\ref{fig:SingleOFDMsymbolNMSExspeed}, we compare the NMSE performance of the interpolated channel within the SRS symbols when estimating the channel using SRS in frequency domain compared to in the in DD domain. We first study the NMSE in function of SNR and consider a fixed speed of v$=500$~km/h, which results in maximum Dopplers of $\upsilon_{\rm{max}}=2130$~Hz. The results in Fig.~\ref{fig:SingleOFDMsymbolNMSE} show that the NMSE performance is significanty improved when the channel is estimated in the DD domain, when compared to the frequency domain estimation at high SNRs. However, at low SNR regimes, the channel estimation in frequency domain has a better NMSE performance. This is due to the larger noise interference while estimating the channel in DD domain, as shown in \eqref{eq:estDDdatainterf}. In Fig.~\ref{fig:SingleOFDMsymbolNMSExspeed} we compare the NMSE performances as a function of maximum speed present in the channel with a fixed SNR of 20~dB. We consider speeds of up to 1000~km/h, in line with the target mobility support of IMT-2030. The results show that DD channel estimation achieves higher performance at speeds higher than approximately 210~km/h. It is worth noting that with the emergence of FR3 bands, the increase in carrier frequency leads to significantly higher Doppler shifts in the channel. For example, if carrier frequencies around 10 GHz are adopted, the NMSE performance would be similar to the scenarios shown in Fig.~\ref{fig:SingleOFDMsymbolNMSExspeed}, but occurring at half the speed.

To study the BER performance of our proposed method, we consider a scenario where the UE transmits 4 slots. The first two slots contains SRS, as shown in \ref{Fig:Slotstructure}, and the last two slots are filled entirely with data. In Fig.~\ref{fig:BER}, we analyze the BER performance of the first three OFDM symbols of the second slot, i.e., symbol indices 15, 16, and 17. We compare the performance of three methods. In the first, the channel is estimated in the frequency domain using two SRS slots, followed by interpolation for the considered data OFDM symbols. In the second and third methods, the channel is estimated in the delay-Doppler domain, being predicted using only the BEM in the second case and using both the BEM and the data symbols as virtual pilots in the third. The signal is transmitted through a channel with maximum speeds of 360~km/h. The results show that the BER of frequency-domain channel estimation reaches a high floor as early as the first symbol and degrades significantly thereafter, failing to reliably estimate data from the second symbol onward. In contrast, the BER of DD domain estimation exhibits slower performance degradation across subsequent symbols and does not encounter a BER floor within the observed interval. Interestingly, in the third method, since we improve the channel estimates using the detected data, the BER performance improves instead of degrading from OFDM symbol 15 to 17.

\begin{figure}
    \centering
    \includegraphics[width=0.9\linewidth]{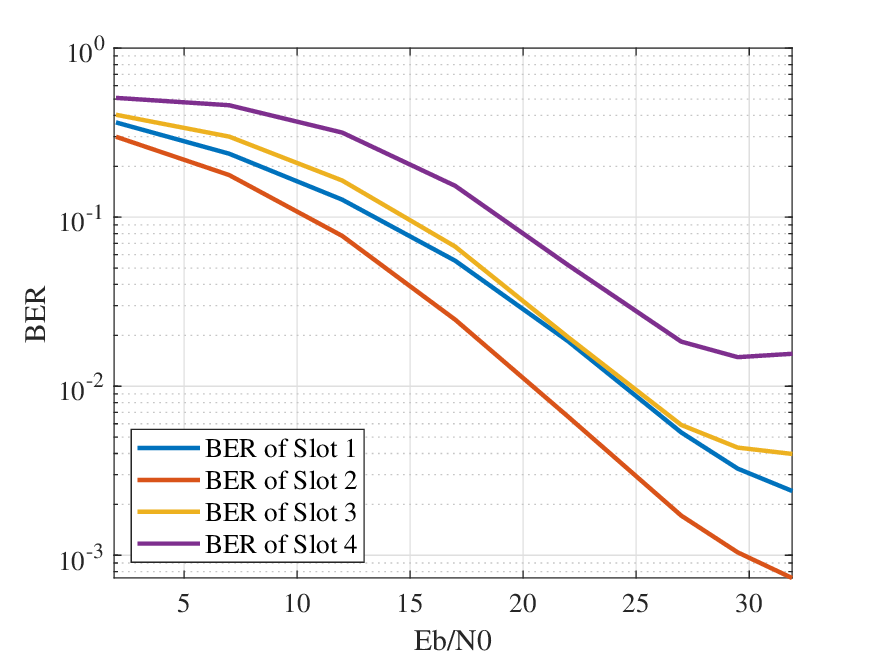}
    \caption{Average BER performance for each transmitted slot.}
    \label{fig:BER3symbdatadriven}
\end{figure}

\begin{figure}
    \centering
    \includegraphics[width=0.9\linewidth]{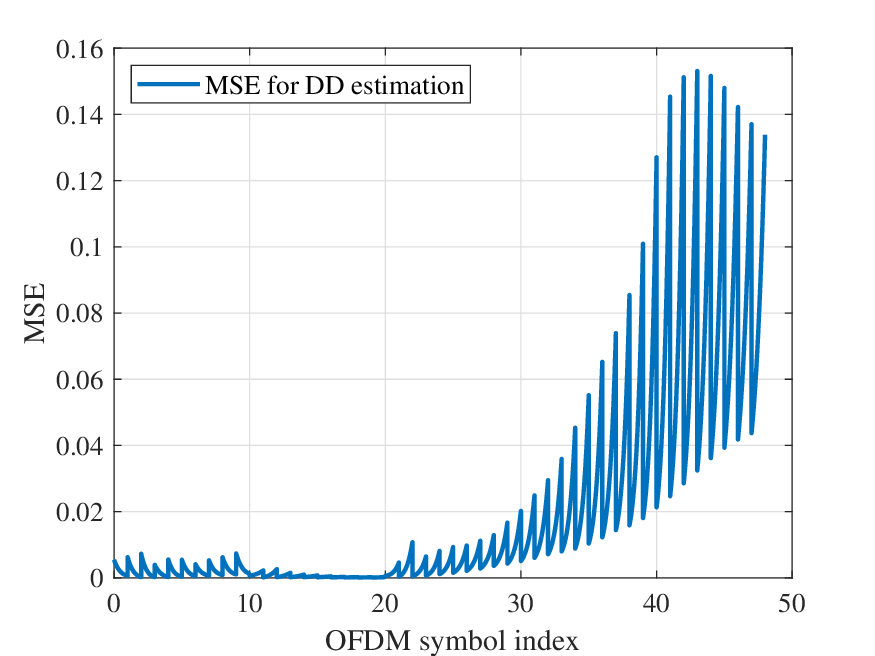}
    \caption{MSE performance in function of OFDM symbol index for the proposed data-driven technique.}
    \label{fig:BERbyslot}
\end{figure}

Finally, in Fig.~\ref{fig:BER3symbdatadriven} and Fig.~\ref{fig:BERbyslot} we study the performance of our proposed method using data as virtual pilots for all the transmitted slots, especially the last two, where only data is present. 
In Fig.~\ref{fig:BER3symbdatadriven}, we show the average BER performance for each of the transmitted slots. The results show an improved BER performance for the data transmitted slot 2, while slot 4 still achieves comparable performance of the OFDM symbol 17 using only the BEM in Fig~\ref{fig:BER}, even after two entire slots without pilot transmission. Fig.~\ref{fig:BERbyslot} shows the MSE of the channel estimate as a function of the OFDM symbol index (sample index/$M_{\rm{T}}$), at SNR of 30~dB. The results show that even though the channel estimates greatly degrades within each OFDM symbol, our proposed method effectively mitigates the channel aging effect and reduces the MSE before detecting data in the next OFDM symbol.

\section{Conclusion}
\label{sec:conc}

In this paper, we introduced a framework that can effectively leverage the advantages of DD domain processing under LTV channels using the 5G NR standard pilots. We employ a DFT-s-OFDM demodulator to transform each OFDM symbol into the DD domain, where the DD channel is estimated. Additionally, we proposed a linear joint channel estimation and equalization technique that sequentially refines the channel estimates and detects the data from subsequent OFDM symbols. 
Through a detailed theoretical analysis, we showed that this framework significantly improves the BER and NMSE performance compared to conventional frequency-domain channel estimation. Furthermore, the results indicate that our proposed method can reliably detect data from transmitted slots without pilot sequence, after an initial channel estimation based on two transmission slots containing SRS. 

\section{Acknowledgment}

This publication has emanated from research supported in part by a research grant from Taighde Éireann (Research Ireland) under Grant Numbers 21/US/3757 and 13/RC/2077\_P2.

\bibliographystyle{IEEEtran}

\bibliography{main}

\begin{thebibliography}{10}
\providecommand{\url}[1]{#1}
\csname url@samestyle\endcsname
\providecommand{\newblock}{\relax}
\providecommand{\bibinfo}[2]{#2}
\providecommand{\BIBentrySTDinterwordspacing}{\spaceskip=0pt\relax}
\providecommand{\BIBentryALTinterwordstretchfactor}{4}
\providecommand{\BIBentryALTinterwordspacing}{\spaceskip=\fontdimen2\font plus
\BIBentryALTinterwordstretchfactor\fontdimen3\font minus \fontdimen4\font\relax}
\providecommand{\BIBforeignlanguage}[2]{{%
\expandafter\ifx\csname l@#1\endcsname\relax
\typeout{** WARNING: IEEEtran.bst: No hyphenation pattern has been}%
\typeout{** loaded for the language `#1'. Using the pattern for}%
\typeout{** the default language instead.}%
\else
\language=\csname l@#1\endcsname
\fi
#2}}
\providecommand{\BIBdecl}{\relax}
\BIBdecl

\bibitem{CientificChallenges6G}
M.~Chafii, L.~Bariah, S.~Muhaidat, and M.~Debbah, ``Twelve scientific challenges for {6G}: Rethinking the foundations of communications theory,'' \emph{IEEE Communications Surveys \& Tutorials}, vol.~25, no.~2, pp. 868--904, 2023.

\bibitem{CodedOTFS}
S.~Li, J.~Yuan, W.~Yuan, Z.~Wei, B.~Bai, and D.~W.~K. Ng, ``Performance analysis of coded {OTFS} systems over high-mobility channels,'' \emph{IEEE Transactions on Wireless Communications}, vol.~20, no.~9, pp. 6033--6048, 2021.

\bibitem{DDaided}
\BIBentryALTinterwordspacing
Y.~Ma, B.~Ai, J.~Yuan, S.~Li, Q.~Cheng, Z.~Shi, W.~Yuan, Z.~Wei, A.~Shafie, G.~Ma, Y.~Lu, M.~Yang, and Z.~Zhong, ``Delay-doppler domain signal processing aided {OFDM (DD-a-OFDM)} for {6G} and beyond,'' 2025. [Online]. Available: \url{https://arxiv.org/abs/2508.04253}
\BIBentrySTDinterwordspacing

\bibitem{MP1}
P.~Raviteja, K.~T. Phan, Y.~Hong, and E.~Viterbo, ``Interference cancellation and iterative detection for orthogonal time frequency space modulation,'' \emph{IEEE Transactions on Wireless Communications}, vol.~17, no.~10, pp. 6501--6515, 2018.

\bibitem{MP2}
F.~Liu, Z.~Yuan, Q.~Guo, Z.~Wang, and P.~Sun, ``Message passing-based structured sparse signal recovery for estimation of {OTFS} channels with fractional doppler shifts,'' \emph{IEEE Transactions on Wireless Communications}, vol.~20, no.~12, pp. 7773--7785, 2021.

\bibitem{BEM2}
Y.~Liu, Y.~L. Guan, and D.~G. G., ``Near-optimal {BEM} {OTFS} receiver with low pilot overhead for high-mobility communications,'' \emph{IEEE Trans. Commun.}, vol.~70, no.~5, pp. 3392--3406, 2022.

\bibitem{SC-FDMA}
A.~Farhang and M.~Bayat, ``{SC-FDMA} as a delay-doppler domain modulation technique,'' in \emph{2024 IEEE International Conference on Communications Workshops (ICC Workshops)}, 2024, pp. 69--74.

\bibitem{Embedded}
P.~Raviteja, K.~T. Phan, and Y.~Hong, ``Embedded pilot-aided channel estimation for {OTFS} in delay–doppler channels,'' \emph{IEEE Transactions on Vehicular Technology}, vol.~68, no.~5, pp. 4906--4917, 2019.

\bibitem{ReducedCP}
P.~Raviteja, Y.~Hong, E.~Viterbo, and E.~Biglieri, ``Practical pulse-shaping waveforms for reduced-cyclic-prefix {OTFS},'' \emph{IEEE Transactions on Vehicular Technology}, vol.~68, no.~1, pp. 957--961, 2019.

\bibitem{BEM1}
H.~Şenol and C.~Tepedelenlioğlu, ``Subspace-based estimation of rapidly varying mobile channels for {OFDM} systems,'' \emph{IEEE Transactions on Signal Processing}, vol.~69, pp. 385--400, 2021.

\bibitem{3gpp_ts38211_v1860}
``{{5G; NR}; Physical channels and modulation} ({3GPP TS} 38.211 version 18.6.0 release 18),'' 3GPP, Tech. Rep. TS 38.211, Apr. 2025, release 18.

\bibitem{3gppTR38901}
``{{5G}; Study on channel model for frequencies from 0.5 to 100 GHz ({3GPP TR} 38.901 version 16.1.0 Release 16)},'' 3rd Generation Partnership Project (3GPP), Technical Report TR 38.901, November 2020, version 16.1.0, Release 16.

\end{thebibliography}

\end{document}